\documentclass[]{aa}  

\usepackage{graphicx}
\usepackage{txfonts}
\usepackage{lipsum}
\usepackage{subcaption}        
\usepackage{lscape}         
\usepackage{placeins}
\usepackage[flushleft]{threeparttable}

\usepackage{hyperref}
\hypersetup{
    colorlinks=true,
    linkcolor=blue,
    citecolor=blue,
    filecolor=blue,      
    urlcolor=blue,
    breaklinks=true,
}
\begin{document}

\newcommand{\Pic}{$\beta$~Pic\xspace}
\newcommand{\water}{H$_2$O\xspace}
\newcommand{\ammonia}{NH$_3$\xspace}
\newcommand{\methane}{CH$_{4}$\xspace}

\title{Signatures of \methane and \ammonia in the atmosphere of $\beta$~Pictoris~b}

\subtitle{}

\author{A.~Barone\inst{\ref{inst:1}, \ref{inst:2}} \corrauth{andrea.barone@astro.su.se}
    \and 
    M.~Janson\inst{\ref{inst:1}}
    \and 
    A.~Brandeker\inst{\ref{inst:1}}
    \and 
    T.~Stolker\inst{\ref{inst:3}}
    \and 
    A.~Vigan\inst{\ref{inst:5}}
    \and 
    K.~Hoy\inst{\ref{inst:6}}
    \and 
    P.~Molliere\inst{\ref{inst:7}}
    \and 
    A.~Denis\inst{\ref{inst:5}}
    \and 
    H.~Anwand-Heerwart\inst{\ref{inst:8}}
    \and
    Y.~Charles\inst{\ref{inst:5}}
    \and
    J.~C.~Costes\inst{\ref{inst:9}}
    \and
    A.~Costille\inst{\ref{inst:5}}
    \and
    M.~El Morsy\inst{\ref{inst:10}}
    \and
    J.~Garcia\inst{\ref{inst:5}}
    \and
    M.~Houllé\inst{\ref{inst:11}}
    \and
    M.~Lopez\inst{\ref{inst:5}}
    \and
    G.~Murray\inst{\ref{inst:12}}
    \and
    E.~Muslimov\inst{\ref{inst:13}}
    \and
    G.~P.~P.~L.~Otten\inst{\ref{inst:4}}
    \and
    J.~Paufique\inst{\ref{inst:14}}
    \and
    M.~Phillips\inst{\ref{inst:15}}
    \and
    U.~Seemann\inst{\ref{inst:14}}
    \and
    A.~Viret\inst{\ref{inst:5}}
    \and
    G.~Zins\inst{\ref{inst:16}}
    }

\institute{
    \label{inst:1} Stockholm University, Astronomy Department, AlbaNova University Center SE-106 91 Stockholm, Sweden
    \and
    \label{inst:2} Dipartimento di Fisica, Università degli Studi di Torino, via Pietro Giuria 1, I-10125, Torino, Italy
    \and
    \label{inst:3} Leiden Observatory, Leiden University, Einsteinweg 55, 2333 CC Leiden, The Netherlands
    \and
    \label{inst:4} Academia Sinica, Institute of Astronomy and Astrophysics, 11F Astronomy-Mathematics Building, NTU/AS campus, No. 1, Section 4, Roosevelt Rd., Taipei 10617, Taiwan
    \and
    \label{inst:5} Aix Marseille Univ, CNRS, CNES, LAM, Marseille, France
    \and
    \label{inst:6} Instituto de Estudios Astrof{\'i}sicos, Facultad de Ingenier{\'i}a y Ciencias, Universidad Diego Portales, Av. Ej{\'e}rcito Libertador 441, Santiago, Chile.
    \and
    \label{inst:7} Max-Planck-Institut für Astronomie, Königstuhl 17, 69117 Heidelberg, Germany
    \and
    \label{inst:8} Institute for Astrophysics, Georg-August University, Friedrich-Hund-Platz 1, 37077 Göttingen, Germany
    \and
    \label{inst:9} Astrophysics Research Centre, School of Mathematics and Physics, Queen’s University Belfast, Belfast BT7 1NN, UK
    \and
    \label{inst:10} Space Telescope Science Institute, 3700 San Martin Drive, Baltimore, MD 21218, USA
    \and
    \label{inst:11} Université Côte d'Azur, Observatoire de la Côte d'Azur, CNRS, Laboratoire Lagrange, France
    \and
    \label{inst:12} Center for Advanced Instrumentation, Durham University, Durham, DH1 3LE, UK
    \and
    \label{inst:13} Department of Physics, University of Oxford, Keble Rd, OX14 3RH Oxford, UK
    \and
    \label{inst:14} European Southern Observatory (ESO), Karl-Schwarzschild-Str. 2, 85748 Garching, Germany
    \and
    \label{inst:15} Institute for Astronomy, University of Edinburgh, Royal Observatory, Blackford Hill, Edinburgh, EH9 3HJ, UK
    \and
    \label{inst:16} European Southern Observatory, Alonso de Cordova 3107, Vitacura, Santiago, Chile
    }

\date{Received 21 Oct, 2026}

  \abstract
   {$\beta$~Pictoris is a nearby (19.7 pc), young ($\sim$ 24 Myr), bright ($V = 3.85$) planetary system hosting three directly imaged planets. Previous high-resolution spectroscopy of \Pic~b has detected \water and CO, but not \methane or \ammonia. Combining observations across a broad wavelength range has been shown to maximise sensitivity to such weak molecular signals.}
   {We searched for previously undetected molecular species in the atmosphere of \Pic~b using high-resolution near-infrared spectra.} 
   {We combined 44 CRIRES+ spectral settings covering the H, K, L, and M bands and cross-correlated the extracted planetary spectra with molecular templates. For species with uncertain detectability, we weighted the observations by their molecular information content and data quality, independent of the measured candidate-species cross-correlation functions.}
   {We recover \water and CO at cross-correlation S/N values of approximately 200 and 50, respectively. We additionally identify tentative \methane and \ammonia signals near the expected radial velocity of \Pic~b, with S/N values of approximately 5.3 and 5.7. Equivalent extractions at the same projected separation on the opposite side of the star show no comparable peaks.}
   {We classify the \methane and \ammonia signals as tentative because their cross-correlation S/N values do not fully account for possible residual systematics. If confirmed, these would be the first detections of both species in the atmosphere of \Pic~b.}


\maketitle
\nolinenumbers

\section{Introduction}
High-resolution spectroscopy (HRS) has become a powerful tool for characterising the atmospheres of exoplanets. HRS resolves molecular bands into tens to hundreds of individual lines, and cross-correlating the observed spectra with a model template enhances the signal for a given species. HRS is highly complementary to low-resolution spectroscopy because, in addition to probing abundant atmospheric species, it is sensitive to species present at relatively low abundances. It has been applied to transiting and close-in planets, where the planet's changing radial velocity shifts its lines relative to the stellar and telluric ones \citep[e.g.][]{Greene2016, Sing2018, Kempton2024}, and it can also be combined with high-contrast imaging (HCI) to study giant planets at wide separations ($>10$\,au) that are spatially resolved from their host stars \citep[e.g.][]{Snellen2015, Wang2016, Wang2021b, Janson2025, Denis2025}. This synergy enables robust and precise measurements of elemental abundance ratios, rotational and radial velocities, and orbital trajectories, all of which are key diagnostics for investigating planet formation pathways, interactions with circumplanetary discs, and the dynamical and migration histories of planetary systems \citep[e.g.][]{Konopacky2013, Snellen2014, Maire2021}.

$\beta$~Pictoris is one of the most remarkable and extensively studied planetary systems. It is a nearby \citep[19.7 pc;][]{GaiaCollaboration2016},  young \citep[$24.7^{+0.9}_{-0.6}$ Myr;][]{Luhman2024}, bright ($V=3.85$) A6V star surrounded by three known giant planets discovered through direct imaging or interferometry \citep{Lagrange2010, Lagrange2019, Nowak2020, Gibbs2026, Sutlieff2026}. \Pic~b is the second planet in order of increasing orbital separation from the host star. Its mass is estimated to be 7--12 $M_{\mathrm{Jup}}$, and it has a semi-major axis of $\sim$10\,au \citep{Lacour2021, Brandt2021}, making it a widely separated super-Jupiter. Owing to its suitability for high-contrast HRS, \Pic~b has been the subject of numerous atmospheric studies. Using CRyogenic high-resolution InfraRed Echelle Spectrograph \citep[CRIRES+, $R\sim50\,000$--$100\,000$;][]{Kaeufl2004, Dorn2023} K-band observations, \citet{Landman2024} reported detections of \water\ and CO. A study in the M-band led by \citet{Parker2024} later confirmed the presence of the species that were claimed to have been detected in the K-band, and reported a tentative detection of SiO. \citet{Janson2025} investigated the system in the L and M bands, confirming the presence of \water\ but not that of SiO. Finally, the recent K-band study by \citet{Gonzalez2026} detected $^{13}$CO and measured a $^{12}$CO/$^{13}$CO ratio consistent with that of the local interstellar medium and lower than the solar photospheric value. 

In this paper, we present a panchromatic analysis of \Pic~b with CRIRES+, covering the near-infrared (NIR) wavelength range from $1.5$ to $5.5\,\mu$m. During ESO Period P115, we acquired 17 new observations in the K, L, and M bands and complemented them with additional NIR spectra from previous publications (see Sect.~\ref{sec: Observations and data reduction}). \citet{Janson2025} demonstrated that combining datasets with different spectral settings increases the signal-to-noise ratio (S/N) more than combining datasets with equal wavelength coverage. The newly acquired data, covering several previously unexplored regions of the spectrum, allow us to improve sensitivity to unexplored molecular species (e.g.\ \methane and \ammonia) and strengthen detections of molecules reported in earlier studies (i.e.\ \water and CO).

\section{Observations} \label{sec: Observations and data reduction}
We observed the \Pic planetary system with the CRIRES+ spectrograph at the VLT between 25 September and 1 October 2025 during ESO Period P115 (Programme ID: 115.2820.001; PI: Janson). We acquired 20 different spectral settings, 17 of which provided complete coverage of the K, L, and M bands; whereas the remaining three, covering the H and J bands, are not included in the analysis (see Table~\ref{tab: observations}). Twelve of the spectral settings acquired during ESO Period P115 and used in this analysis cover a wavelength range which was never observed before in HRS.

A fixed instrumental setup was used for all epochs, with a slit width of 0.2$^{\prime \prime}$, yielding a spectral resolving power of $R \sim 100\,000$. The slit was aligned with the edge-on plane of the planetary system (position angle, PA$=31.6^\circ$) to encompass the host star and all the planets, \Pic~b, c and d. We are currently analysing planets c and d and will present the results in a subsequent paper. The observations were performed using the MACAO adaptive optics system \citep{Paufique2004} and an ABBA nodding pattern to enable pairwise subtraction of the high sky background in the L and M bands. For all L-band settings, we adopted a direct integration time (DIT) of 30\,s per exposure, with four exposures per nodding position over 11 nodding cycles. For the M-band settings, the DIT was 5\,s, and the number of detector integrations (NDIT) was 12, with two exposures per nodding position over 11 nodding cycles. The two K-band observing blocks (OBs) used a DIT of 10 s and an NDIT of 12, with two exposures per nodding position over six nodding cycles. The J- and H-band settings acquired during this programme are not included in the present analysis and will instead be used in a dedicated study of the circumstellar disc.

In addition to the newly acquired observations, we included 13 CRIRES+ L- and M-band datasets from \citet[Programme ID: 110.23NC.001; PI: Janson]{Janson2025}, two CRIRES+ K-band datasets from \citet[Programme ID: 108.22HG.001; PI: Stolker]{Landman2024}, one CRIRES+ K-band dataset from \citet[Programme ID: 114.27DX.001; PI: Kenworthy]{Gonzalez2026}, two CRIRES+ M-band datasets from \citet[Programme ID: 109.23G2.001, PI: Birkby]{Parker2024}, two additional CRIRES+ M-band datasets from the CRIMSON survey \citep[Programme ID: 114.27LL.002; PI: Parker;][]{Parker2026}, and seven H-band spectra from the High-Resolution Imaging and Spectroscopy of Exoplanets \citep[HiRISE;][]{Vigan2024} survey (Programme ID: 114.27C6.001; PI: Hoy).

\begin{figure*}
\centering
\sidecaption
   \includegraphics[width=12cm]{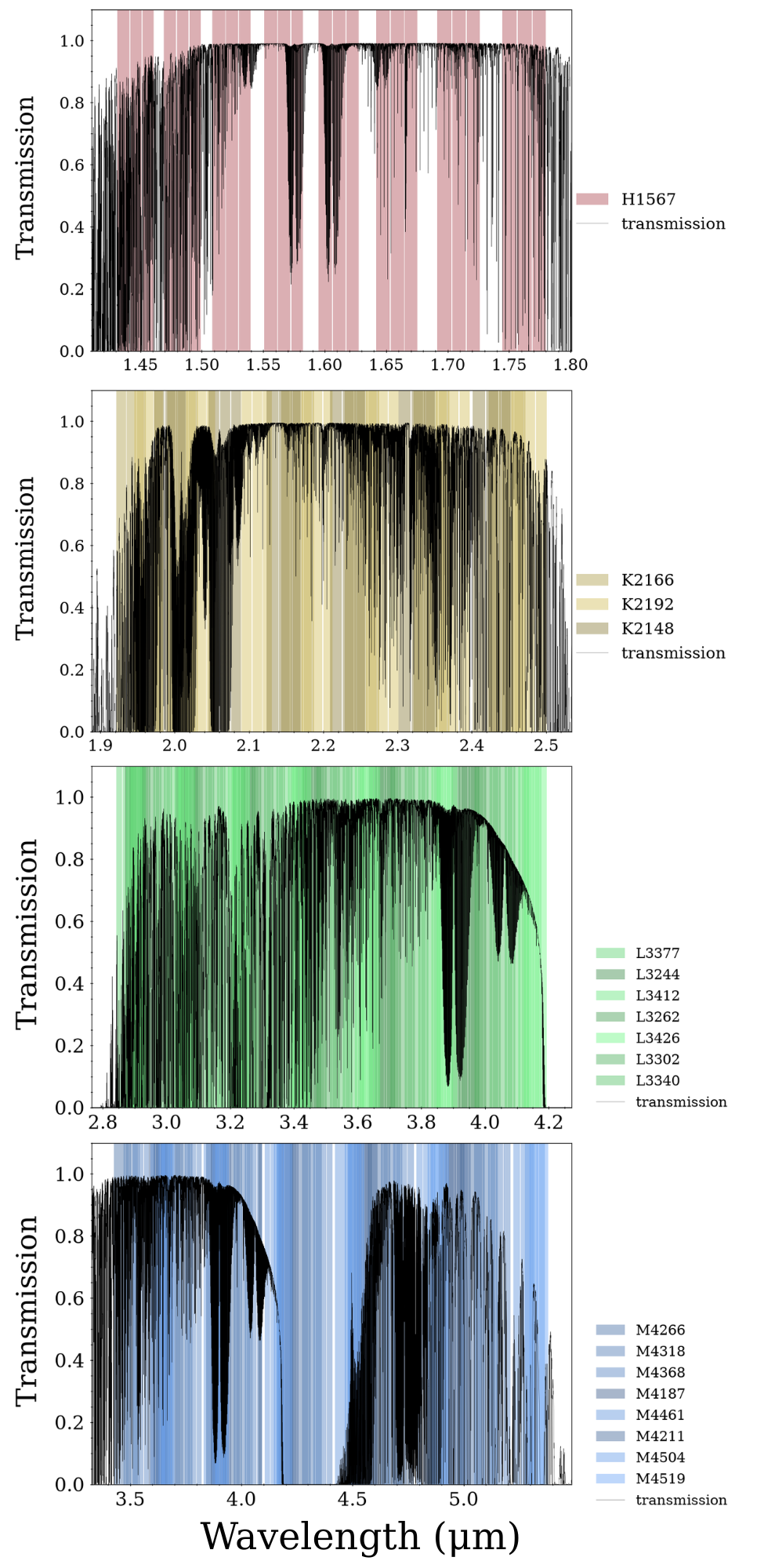}
     \caption[]{From top to bottom: spectral ranges covered by detector orders in the grating settings used for this analysis, divided into H, K, L and M bands, respectively. The observations cover almost all of the four bands. The overplotted black line is the fractional atmospheric transmission. These plots were generated using \texttt{CRIRESViz} \footnotemark}
     \label{fig: Orders}
\end{figure*}

Combining these observations yielded a dataset comprising 44 spectral settings and covering nearly the entire near-infrared wavelength range ($1.5{-}5.5 \, \mu$m). This broad wavelength coverage enables a panchromatic analysis, which has been shown to maximise the S/N \citep[][]{Janson2025}. To our knowledge, this represents the broadest high-resolution wavelength coverage obtained for a directly imaged exoplanet to date (see Fig.~\ref{fig: Orders}).

\begin{table}[]
    \centering
    \caption{\Pic observations acquired with CRIRES+ in September 2025 (Programme ID: 115.2820.001).}
    \begin{tabular}{ccccc}
    \hline
    \hline
        Date & Spectral Setting & DIT & NDIT & $N_{\text{fr}}^{\dag}$ \\ [0.3ex]
        \hline
         25 Sep 2025 & L3262 & 30 & 1 & 86 \\ 
         25 Sep 2025 & L3340 & 30 & 1 & 86 \\ 
         25 Sep 2025 & M4187 & 5 & 12 & 42 \\ 
         27 Sep 2025 & L3426 & 30 & 1 & 86 \\ 
         27 Sep 2025 & M4318 & 5 & 12 & 42 \\ 
         27 Sep 2025 & M4461 & 5 & 12 & 42 \\
         28 Sep 2025 & L3244 & 30 & 1 & 86 \\ 
         28 Sep 2025 & L3302 & 30 & 1 & 86 \\ 
         28 Sep 2025 & M4211 & 5 & 12 & 42 \\
         29 Sep 2025 & L3377 & 30 & 1 & 86 \\ 
         29 Sep 2025 & M4368 & 5 & 12 & 42 \\ 
         29 Sep 2025 & M4504 & 5 & 12 & 42 \\
         30 Sep 2025 & L3412 & 30 & 1 & 86 \\ 
         30 Sep 2025 & K2192 & 10 & 12 & 22 \\ 
         30 Sep 2025 & K2148 & 10 & 12 & 22 \\
         30 Sep 2025 & M4266 & 5 & 12 & 42 \\
         01 Oct 2025 & M4519 & 5 & 12 & 42 \\ 
         01 Oct 2025 & H1567$^*$ & 10 & 12 & 22 \\ 
         01 Oct 2025 & H1582$^*$ & 10 & 12 & 22 \\
         01 Oct 2025 & J1226$^*$ & 10 & 10 & 22 \\
         \hline
         \hline
    \end{tabular}
    \begin{tablenotes}
          \small
          \item $\dag$: Number of frames in the run.
          \item $*$: These data sets were not included in this analysis. 
    \end{tablenotes}
    \label{tab: observations}
\end{table}

\footnotetext{\url{https://github.com/andrew070500/CRIRESViz.git}}

\section{Data reduction} \label{sec: Data reduction} Given the broad wavelength coverage of our dataset, different reduction and extraction strategies were adopted depending on the spectral band. 

\subsection{L- and M-band} \label{ssec: L- and M-band}
We reduced the L- and M-band data with the custom pipeline\footnote{\url{https://github.com/AWehrhahn/CRIRES_LM}} described in \cite{Janson2025}. After bias, dark-current, and flat-field correction, the pipeline derived the slit geometry and obtained the wavelength solution from telluric absorption and emission features.

The point spread function (PSF) subtraction scheme performed here is a variant of the STARK\footnote{\url{https://github.com/Jayshil/stark}} routine previously implemented on VLT/UVES \citep[][]{Ringqvist2023} and JWST/NIRCAM \citep[][]{Patel2024} data. Stellar PSF subtraction was performed on the pairwise background-subtracted images. First, a non-uniform grid was constructed in wavelength (dispersion direction) and spatial separation. A two-dimensional spline was then fitted in wavelength-separation space using a non-uniform grid of knots. The knot positions were chosen in normalised-flux space, ensuring that the spline model adequately samples regions containing both continuum and telluric features. Consequently, this naturally ensures that deep telluric features are represented in the spline model, and these features are removed together with the stellar contribution during the PSF subtraction. This procedure yielded a two-dimensional model of the stellar PSF, which was subsequently subtracted from the data.  

To extract the planetary signal, we derived its position using the \texttt{whereistheplanet} package \citep{Wang2021} and extracted its spectrum using the same spatial PSF profile derived during the construction of the stellar PSF model. We performed the same extraction at the same projected separation on the opposite side of the star, where no planetary signal is expected, to provide a reference spectrum for estimating the noise properties in subsequent analyses. We divided the obtained spectra by the stellar one to obtain the data in contrast space.

\subsection{K-band}  \label{ssec: K-band}
Together with the newly observed K-band data and the data from \citet{Landman2024}, which were all taken with the slit aligned with the system's orbital plane, we also collected a K-band spectrum from \citet{Gonzalez2026}, where the slit was set perpendicularly to the orbital plane to increase the integration time. The two sets of observations have been reduced separately.  

\subsubsection{Slit-aligned data}
We reduced the slit-aligned K-band observations with \texttt{pycrires}\footnote{\url{https://github.com/tomasstolker/pycrires}}, following \cite{Landman2024}. After the standard calibrations and wavelength refinement, we fitted a two-dimensional spline to the stellar PSF. The stellar flux was then measured through a weighted extraction within a 5-pixel aperture centred on the PSF. We then subtracted the stellar and residual systematic signals, and extracted the planetary spectrum using the fitted spatial profile at the predicted planet position. We divided the result by the stellar spectrum to obtain planet-to-star contrast.

\subsubsection{Slit-perpendicular data}
For a complete description of these K-band data, refer to \cite{Gonzalez2026}. Briefly, we used the \texttt{excalibuhr} pipeline \citep[][]{Zhang2024}, which reduced the data using standard calibration, sky-subtraction, and frame-stacking procedures. The planetary spectrum was then extracted using a method optimised for faint companions against a strong slit background, with spatial rows weighted according to their cross-correlation strength.

\subsection{H-band} \label{ssec: H-band}
We reduced the HiRISE observations with the pipeline described by \cite{Costes2024} and \cite{Denis2025}, which combines the official CRIRES+ pipeline with custom routines for frame combination, extraction, filtering, and wavelength recalibration. The stellar and planetary spectra were extracted using a Gaussian-like profile and employing optimal extraction, as described in \cite{Denis2026}. We then divided the planet's spectra by the stellar spectrum to obtain planet-to-star contrast.

\section{Analysis} \label{sec: Analysis}
To analyse \Pic b data, we generated model templates and cross-correlated the reduced spectra with them to sum up and enhance the emission-line signal. In this section, we report the details of the model generation, the cross-correlation and the results. 

\subsection{Model generation} \label{ssec: Model generation}
A detailed description of the model generation is provided in \citet{Janson2025}. We used \texttt{petitRADTRANS} models \citep{Molliere2019, Molliere2020} as templates, adopting the relatively well-constrained physical properties of planet b. The model atmosphere consisted of hydrogen and helium, which contribute to the atmospheric pressure but have no strong lines in the wavelength range considered here, together with each molecule searched for in the extracted spectra. We adopted the temperature-pressure profile presented in \citet{Guillot2010}, assuming a negligible contribution from irradiation. For the \Pic~b model, we set the temperature to $1700$ K, following \citet{Parker2024}, the surface gravity $\log{g}$ to $4.18$, and the rotational velocity to $20 \, \text{km s}^{-1}$. Finally, we fitted and subtracted the continuum from the model spectra to retain only the line information and rotationally broadened them to match the expected planetary spectra.

\subsection{Cross-correlation} \label{ssec: Cross-correlation}
Each extracted spectrum is naturally divided into CRIRES+ spectral orders. Thus, each observation comprises 4--7 spectral chunks, which we cross-correlated with the models described in Sect. \ref{ssec: Model generation}. The model templates span the full wavelength coverage of the data and were masked to match the wavelength range of each chunk. They were then Doppler-shifted to the planet's radial velocity at the time of the observation. This procedure yielded a set of $\sim$250 cross-correlation functions (CCFs) per model. In the presence of a significant signal, the corresponding CCFs are expected to exhibit a peak centred at zero velocity. A coherent set of CCFs is required to combine the individual functions, increase the overall S/N, and recover the often hidden signals of more exotic species. A simple average over all spectral regions is not optimal because it does not account for variations in line density among the spectral settings or differences in atmospheric conditions between observing dates. We therefore divided the merging procedure into two steps.

\begin{figure*}
\centering
\sidecaption
   \includegraphics[width=12.75cm]{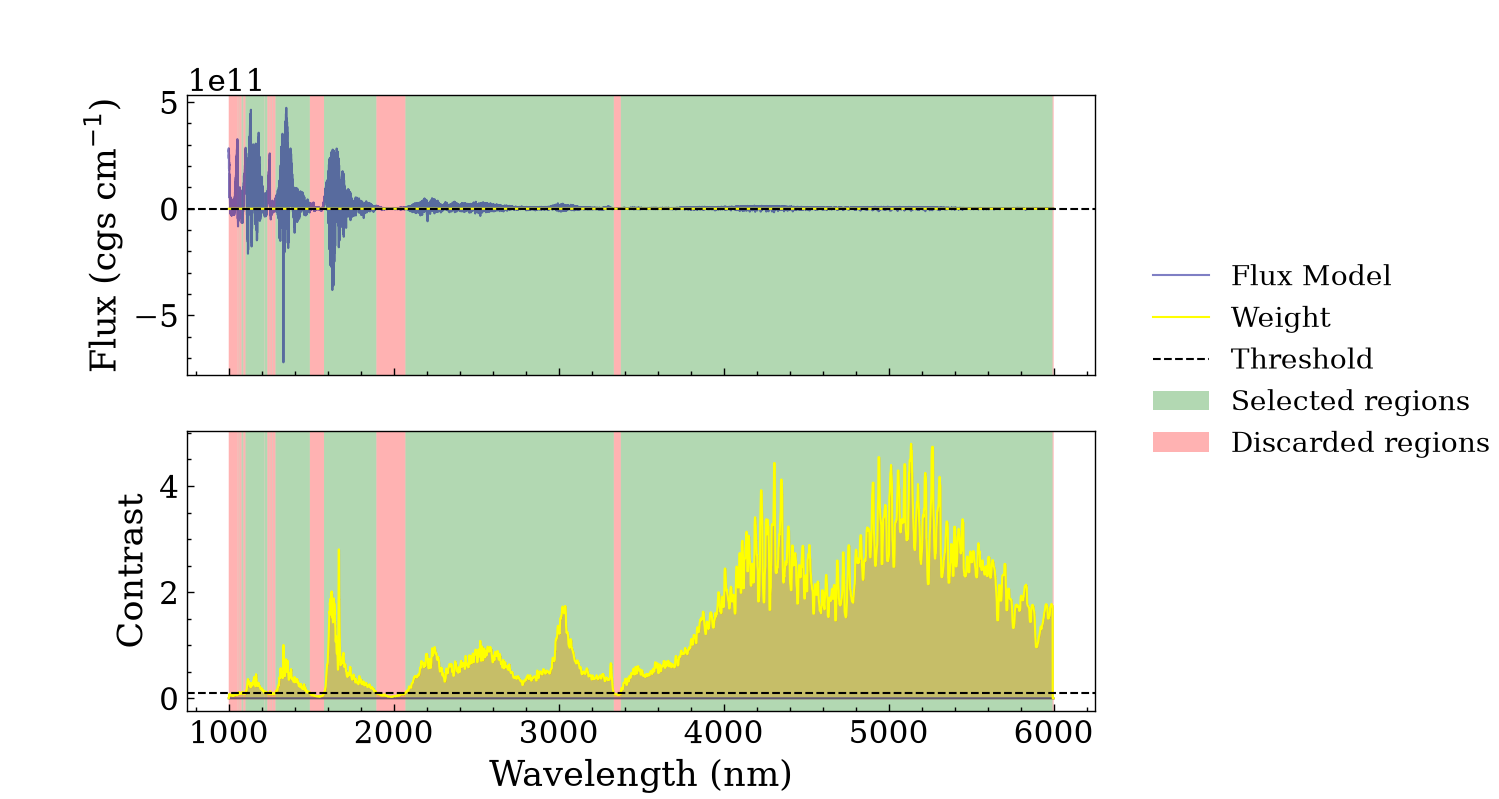}
     \caption{The upper panel shows, in blue, the simulated planetary flux containing \methane and produced using \texttt{petitRadTRANS}. The model is baseline-subtracted. The bottom panel shows the planetary contrast, in yellow, derived from the upper panel and used in the cross-correlation procedure. The black dashed line represents the threshold (i.e. 0.1 in contrast space) used to select the regions containing information. The green areas denote where the model's gradient exceeds the threshold (selected regions), while the red ones are discarded regions.}
     \label{fig: threshold}
\end{figure*}

First, we obtained the CCF for each observation (for example, L3262, acquired on 25 September 2025) by combining the CCFs from its spectral orders (for example, five chunks). In this first step, each chunk was weighted according to its expected information content. To estimate this information content, we defined a threshold (i.e. 0.1) for the model spectral gradient and counted the data points at which the derivative of the model spectrum exceeded this threshold (see Fig. \ref{fig: threshold}). Consequently, orders containing more spectral information contributed more strongly to the CCF of the corresponding spectral setting. This left us with 44 CCFs per model, one per observation. To merge the observations' CCFs, we adopted two different methods depending on the molecule we were searching for. 

For molecular species expected to be detected, such as H$_2$O and CO, we weighted each observation $i$ according to its measured S/N:
\begin{equation} \label{eq: 1}
    S/N_{i} = \frac{\max{(\text{CCF}_i[\text{peak region}])}}{\text{std}{(\text{antiCCF}_i[\text{out-of-peak region}])}}
\end{equation}
Here, the ``peak region'' corresponds to the expected width of the CCF peak, based on the rotational broadening of the planet and the peak biasing \citep[$\sim70\,\mathrm{km\,s^{-1}}$;][]{Parker2024}, while the $\mathrm{antiCCF}_{i}$ ``out-of-peak region'' comprises the complementary range in velocity space, at the same projected separation from the star as the planet but on the opposite side (as in \citealt{Landman2024}), and therefore provides a reference estimate of the systematic noise. Thus, we assigned higher S/N observations greater weight when combining the CCFs. After merging all CCFs, we evaluated the total S/N using Eq. \ref{eq: 1}. 

For molecular species whose detectability was uncertain, such as NH$_3$ and CH$_4$, we adopted a more conservative weighting scheme to avoid artificially enhancing the combined S/N. Using the H$_2$O analysis, we derived a correction factor for each observation:
\begin{equation} \label{eq: 2}
    C_i = S/N_{\text{H$_2$O},i}\cdot\frac{\sigma_{\text{H$_2$O},i}}{\sum{\text{info}_{\text{H$_2$O},i}}},
\end{equation}
where $\sigma_{\mathrm{H_2O},i}$ is the median of the pixel-by-pixel uncertainties and $\sum \mathrm{info}_{\mathrm{H_2O},i}$ is the total information content of the observation's wavelength range, as defined above. The correction factor therefore corresponds to the ratio between the measured H$_2$O S/N and that expected from the information content and estimated uncertainties of each observation.
For a new molecular species, we multiplied $C_i$ by the ratio between its information content and estimated uncertainty:
\begin{equation} \label{eq: 3}
w_{X,i} = C_{i}\cdot\frac{\sum{\text{info}_{X,i}}}{\sigma_{X,i}},
\end{equation}
where $X$ denotes the molecule of interest. The resulting quantity provides an observation-specific estimate of the expected S/N and was used as the weight when combining the CCFs. This approach assigns the weights independently of the measured CCF of the molecule being searched for, thereby reducing the risk of artificially increasing the significance of a noise fluctuation. We adopted H$_2$O as the reference species because of its exceptionally strong detection in our data (see Sect.~\ref{ssec: Results}). When searching for molecular species whose detectability was uncertain, we found a higher noise level at the planetary position than at the corresponding position on the opposite side of the star. This excess may arise from residual systematics introduced by the model's imperfections or by degeneracies within the molecular lines. We emphasise that the model was constructed specifically for molecular detection and is not intended to provide a physically or chemically self-consistent representation of the planetary atmosphere, which will be the goal of an upcoming in-depth analysis. We therefore adopted a more conservative noise estimate. We autocorrelated the model, normalised the resulting autocorrelation function to the observed CCF, and subtracted it from the observed CCF. For \methane and \ammonia, we therefore replaced the denominator of Eq. \ref{eq: 1} with the standard deviation of the residuals obtained after subtracting the scaled model autocorrelation, excluding the peak region. 

\begin{figure*}[]
    \centering        
    {\includegraphics[width=0.45\linewidth, trim={1cm 2.4cm 0.5cm 2.4cm}, clip]{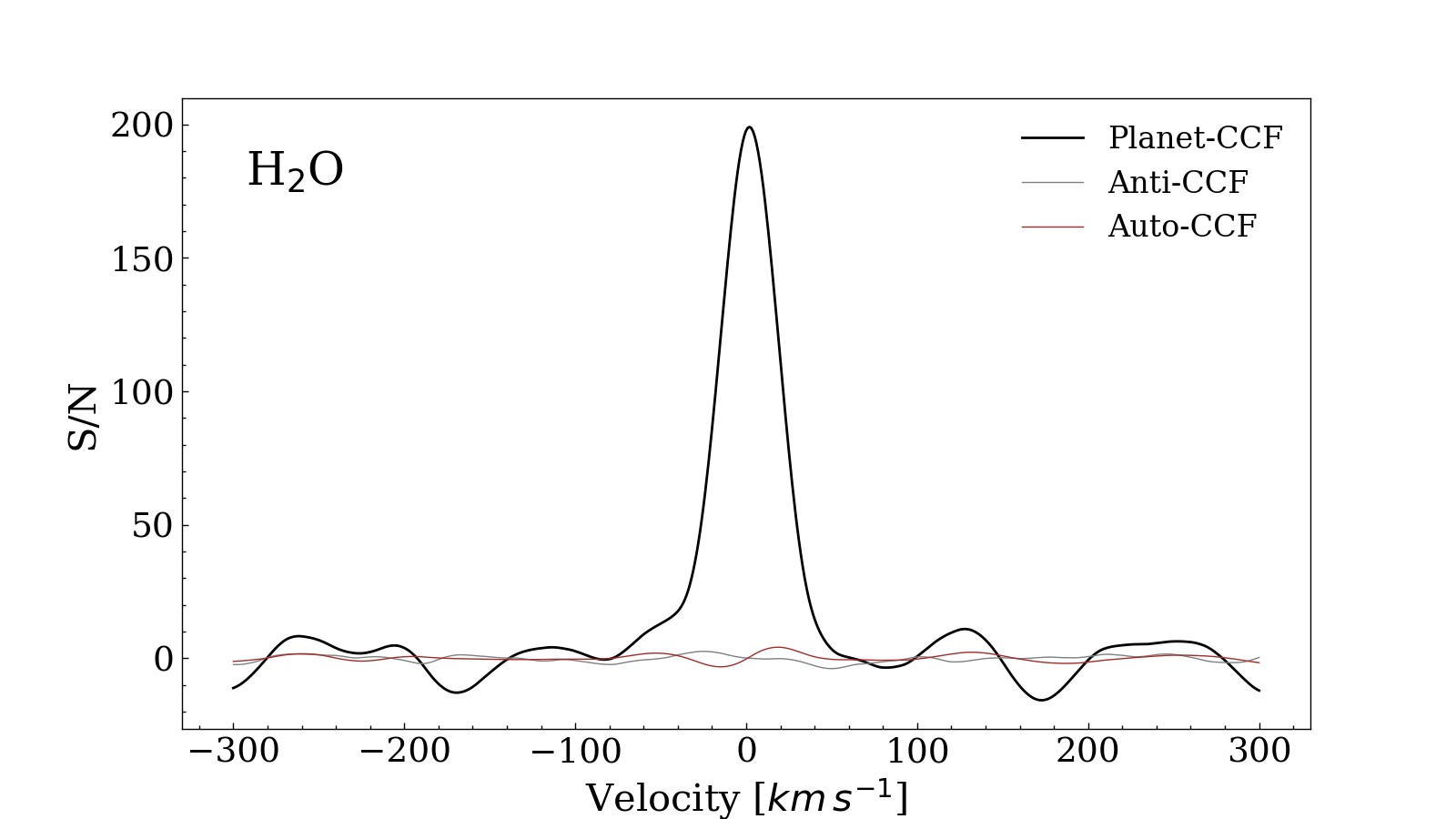}} 
    \hspace{-25pt}
    {\includegraphics[width=0.45\linewidth, trim={1cm 2.4cm 0.5cm 2.4cm}, clip]{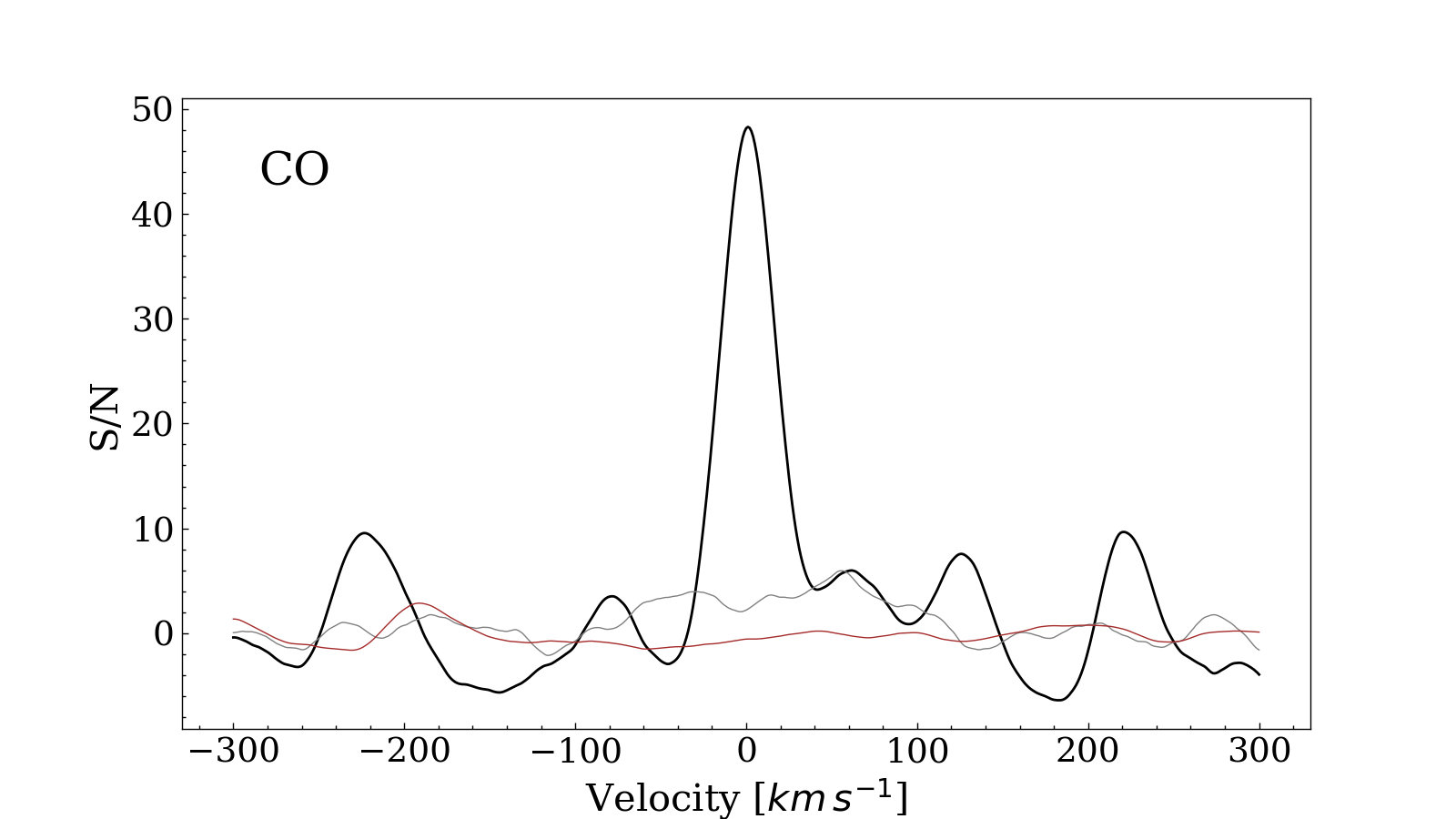}} \\
    {\includegraphics[width=0.45\linewidth, , trim={1cm 0 0.5cm 2.6cm}, clip]{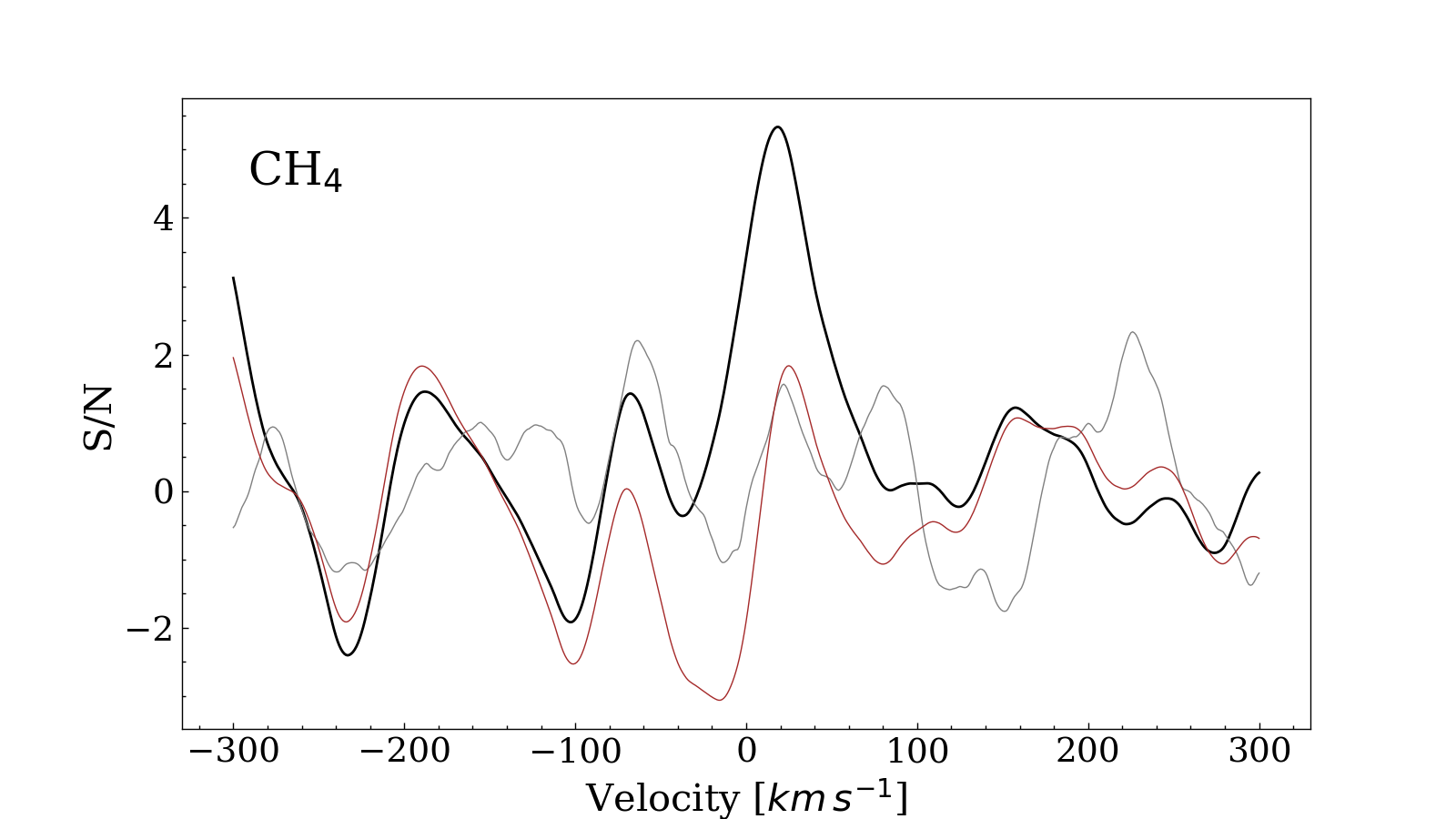}} 
    \hspace{-25pt}
    {\includegraphics[width=0.45\linewidth, , trim={1cm 0 0.5cm 2.6cm}, clip]{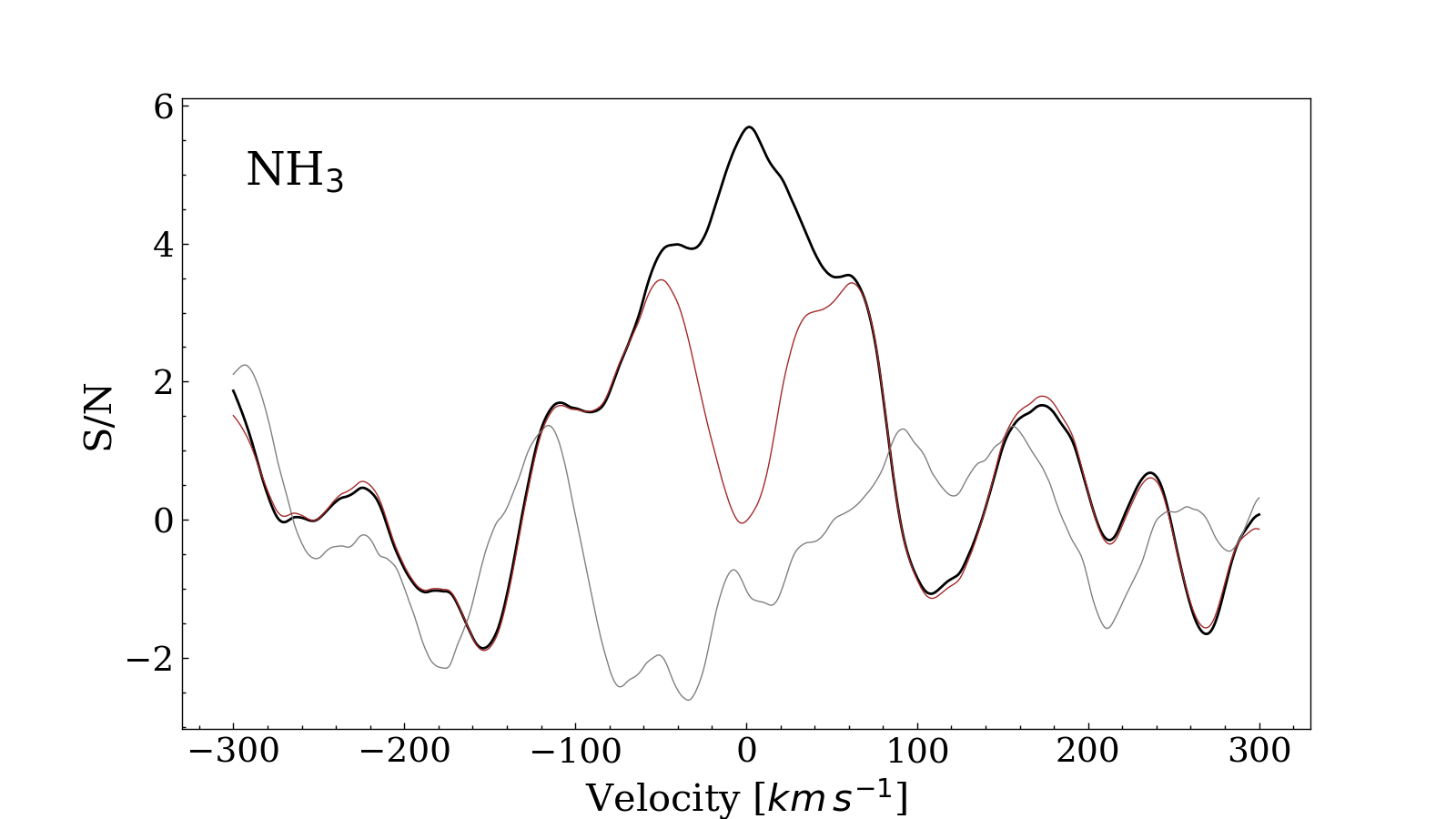}}
    \caption{Cross-correlation functions of \Pic~b with an \water (upper-left panel), CO (upper-right panel), \methane (bottom-left panel), and \ammonia (bottom-right panel). The black lines are the cross-correlation functions at \Pic~b's location; the grey lines are the corresponding functions at the same projected separation on the opposite side of the star; meanwhile, the brown lines show the residuals after subtraction of the scaled model autocorrelation from the planetary CCF.  
    } \label{fig: CCFs}
\end{figure*}

\subsection{Results} \label{ssec: Results}
The analysis yielded strong molecular signals. Consistent with previous studies \citep[e.g.][]{Landman2024, Janson2025}, \water produced the strongest signal. Thanks to the broad wavelength coverage of our dataset, we detected \water at an S/N of $\sim200$ (upper-left panel of Fig.~\ref{fig: CCFs}). The HiRISE H-band data provide the largest contribution, accounting for most of the total signal. Nevertheless, the K-, L-, and M-band data alone yield an S/N of $\sim50$, demonstrating the substantial gain obtained by combining datasets covering different spectral settings rather than repeatedly observing the same settings. Similarly, we detected CO at an S/N of $\sim50$ (upper-right panel of Fig.~\ref{fig: CCFs}). For this molecule, the slit-perpendicular K-band data contributed the most to the final S/N, also more than the slit-aligned data at the same wavelengths. This finding is in agreement with what is presented in \citet{Gonzalez2026}, which demonstrated that orienting the slit perpendicular to the orbital plane, using it as a pseudo-coronagraph, allows for a higher integration time, resulting in higher S/N. 

Before searching for additional molecular species, we removed the contribution of \water from the H-band data because its strong spectral features affected the cross-correlation results for less abundant species, including by generating spurious CCF peaks. We removed the \water contribution using an iterative subtraction procedure. Specifically, from each H-band chunk (as defined in Sect. \ref{ssec: Cross-correlation}), we subtracted a scaled, normalised \water model shifted to the expected planetary radial velocity for that observation and repeated the cross-correlation analysis, varying the scaling factor to minimise the residual \water CCF signal. The resulting H-band residuals, from which the dominant \water contribution had been removed, improved the sensitivity of our searches for molecular species other than \water and CO. 

We searched for a broad range of molecular species (e.g.\ CO$_2$, SiO, PH$_3$, HCN, O$_3$, etc.), including both those expected and those not expected to be present, to test our detection framework. Most species were not detected, with the exception of \methane and \ammonia. We identify candidate \methane and \ammonia signals with CCF S/N values of approximately 5.3 and 5.7, respectively (bottom-left and bottom-right panels of Fig.~\ref{fig: CCFs}). The \ammonia CCF peaks at the expected velocity of \Pic~b, while the \methane peak is offset by approximately $15\,\mathrm{km\,s^{-1}}$, which is small compared with the width of the CCF peak ($\sim70\,\mathrm{km\,s^{-1}}$). Applying the same weighting and combination procedure to spectra extracted at the corresponding position on the opposite side of the star produces no comparable peak, with absolute CCF S/N values below 2.6 for \methane and 1.6 for \ammonia. Before \water subtraction, the \methane CCF contained an additional spurious peak near $80\,\text{km s}^{-1}$, which we traced to the strong H-band \water signal. This emphasises the importance of the water-subtraction procedure. The HiRISE H-band and slit-perpendicular K-band observations received the largest weights and are primarily responsible for the detection of both species. This is due to the significant difference in S/N between the acquisition methods (i.e. slit-aligned spectroscopy vs fibre-fed/slit-perpendicular spectroscopy), showing that, when looking for lower-abundance species, it may be preferable to prioritise the S/N (e.g. observing with HiRISE) more than observing at the wavelength of the fundamental band (e.g. $3.3 \, \mu$m for \methane). However, the contrast with the L- and M-band weights is less pronounced than for \water and CO. This highlights the strength of a multi-band analysis when searching for weak molecular signals. To our knowledge, these are the first reported candidate \methane and \ammonia signals in the atmosphere of \Pic~b, and one of the first tentative detections of these species using HRS in an exoplanet's atmosphere. Under the assumption of chemical equilibrium, theoretical models \citep[e.g.][]{Burrows1997, Sudarsky2000, Lodders2002} predict that, at the pressure and temperature expected for \Pic~b, the atmosphere should be dominated by H$_2$ and He, with \water, CO, and N$_2$ as the main molecular species. This scenario is consistent with the high-significance detections of \water\ and CO, while also allowing \ammonia\ and \methane\ to be present at lower abundances. Non-equilibrium chemistry models predict low \methane and \ammonia abundances in low-surface-gravity planets. However, the high mass, temperature and pressure of \Pic~b are comparable to those of some brown dwarfs in which \methane and \ammonia are predicted to be present \citep[e.g.][]{Zahnle2014, Phillips2020}. Moreover, methane has also been detected in some of these objects. For instance, in the atmospheres of PSO318 \citep{Liu2013} and LRL 11001, which have a similar mass to \Pic~b, \citet{Beiler2023_2} and \citet{Luhman2026} found \methane absorptions and \methane emissions, respectively. Furthermore, a \methane detection would also be consistent with recent findings in the atmosphere of self-luminous exoplanets. For example, \citet{Denis2026} found methane signatures in 51 Eri b, a planet in the \Pic moving group (BPMG) with similar mass and age to those of \Pic~b.

Although both signals have CCF S/N values exceeding 5, we classify them as tentative because this statistic does not fully account for possible residual systematics. A more comprehensive analysis, including atmospheric retrievals and additional tests of the robustness of the signals, is currently underway and will be used to assess whether the detections can be confirmed.

\section{Conclusions} \label{sec: Conclusions}
We combined 44 CRIRES+ settings spanning the H to M bands to conduct a panchromatic molecular search in \Pic~b. We recover \water and CO at CCF S/N values of approximately 200 and 50 and identify tentative \methane and \ammonia signals near the expected planetary velocity, with CCF S/N values of 5.3 and 5.7. Equivalent extractions on the opposite side of the star show no comparable peaks. These results demonstrate the molecular sensitivity gained by combining high-resolution observations across multiple near-infrared bands. A low/medium-resolution analysis using, for instance, JWST/NIRSpec \citep{Jakobsen2022} or MIRI \citep{Rieke2015}, would probably not be sensitive enough to detect lower-abundance species, such as \methane and \ammonia. However, a low-resolution observational campaign might still be worthwhile to cover a larger wavelength range and complement the high-resolution observations. \methane and \ammonia's detections are in agreement with theoretical models and recent detections in brown dwarfs and self-luminous exoplanets. A forthcoming analysis will investigate the robustness and atmospheric implications of the candidate signals using additional validation tests and physically self-consistent atmospheric modelling.

\section{Data availability}
The reduced spectra, together with the models used for the cross-correlation, and the CCFs shown in Fig.~\ref{fig: CCFs} are only available in electronic form at the CDS via anonymous ftp to \url{cdsarc.u-strasbg.fr} (130.79.128.5) or \url{via http://cdsweb.u-strasbg.fr/cgi-bin/qcat?J/A+A/}.

\begin{acknowledgements}
We thank the referee for the comments and suggestions that helped us to improve the quality of our paper. The authors thank D.~Gonz{\'a}lez~Picos for lending part of the K-band data. A.Ba. thanks L.~Parker for the insightful discussion. This project has received funding from \emph{Agence Nationale de la Recherche} (ANR) under grant ANR-23-CE31-0006-01 (MIRAGES). The HiRISE instrument has been developed with funding from the European Research Council (ERC) under the European Union's Horizon 2020 research and innovation programme, grant agreements No. 757561 (HiRISE) and 678777 (ICARUS), from the \emph{Commission Spécialisée Astronomie-Astrophysique} (CSAA) of CNRS/INSU, and from the \emph{Action Spécifique Haute Résolution Angulaire} (ASHRA) of CNRS/INSU co-funded by CNES, and from Région Provence-Alpes-Côte d'Azur under grant agreement 2014-0276 (ASOREX).
\end{acknowledgements}

\bibliographystyle{aa}
\bibliography{Bibliography}

\begin{appendix}

\end{appendix}
\end{document}